\documentclass[review,number,sort&compress]{elsarticle}

\usepackage{tikz}
\usepackage{float}
\usepackage{graphicx}
\usepackage{gensymb}
\usepackage{amssymb}
\usepackage{color,soul}
\usepackage{url}
\usepackage{ulem}
\usepackage{textcomp}
\usepackage[figuresright]{rotating}
\usepackage{subcaption}
\usepackage{subcaption}
\usepackage{hyperref}
\usepackage{cleveref}
\usepackage[printwatermark]{xwatermark}

\definecolor{orange}{rgb}{1,0.5,0}
\definecolor{cadmiumgreen}{rgb}{0.0, 0.42, 0.24}

\DeclareGraphicsExtensions{.eps,.pdf,.png,.jpg,}

\journal{Nuclear Instruments and Methods A}

\begin{document}

\newcommand{\etal}{{\it et~al.}}
\newcommand{\geant} {{{G}\texttt{\scriptsize{EANT}}4}}

\begin{frontmatter}


\title{Response of a Li-glass/multi-anode photomultiplier detector to $\alpha$-particles from $^{241}$Am}


\author[lund]{E.~Rofors}
\author[lund,ess]{H.~Perrey}
\author[ess,glasgow]{R.~Al~Jebali}
\author[glasgow]{J.R.M.~Annand}
\author[glasgow]{L.~Boyd}
\author[julichZ]{U.~Clemens}
\author[LLB]{S.~Desert}
\author[julich]{R.~Engels}
\author[lund,ess]{K.G.~Fissum\corref{cor1}}
\ead{kevin.fissum@nuclear.lu.se}
\author[julich]{H.~Frielinghaus}
\author[ideas]{C.~Gheorghe}
\author[ess,midswe]{R.~Hall-Wilton}
\author[julich]{S.~Jaksch}
\author[lund]{A.~Jalg\'{e}n}
\author[ess]{K.~Kanaki}
\author[julich]{G.~Kemmerling}
\author[lund]{V.~Maulerova}
\author[lund]{N.~Mauritzson}
\author[glasgow]{R.~Montgomery}
\author[lund,ess]{J.~Scherzinger\fnref{fn2}}
\author[glasgow]{B.~Seitz}

\address[lund]{Division of Nuclear Physics, Lund University, SE-221 00 Lund, Sweden}
\address[ess]{Detector Group, European Spallation Source ERIC, SE-221 00 Lund, Sweden}
\address[glasgow]{SUPA School of Physics and Astronomy, University of Glasgow, Glasgow G12 8QQ, Scotland, UK}
\address[midswe]{Mid-Sweden University, SE-851 70 Sundsvall, Sweden}
\address[LLB]{LLB, CEA, CNRS, Universit\'e Paris-Saclay, 91191 Gif-sur-Yvette, France}
\address[julich]{J\"ulich Centre for Neutron Science JCNS, Forschungszentrum J\"ulich, D-52425 J\"ulich, Germany}
\address[julichZ]{Zentrum f\"ur Anwendungsentwicklung und Elektronik ZEA-2, Forschungszentrum J\"ulich, D-52425 J\"ulich, Germany}
\address[ideas]{Integrated Detector Electonics AS, Gjerdrums Vei 19, N-0484 Oslo, Norway}

\cortext[cor1]{Corresponding author. Telephone:  +46 46 222 9677; Fax:  +46 46 222 4709}
\fntext[fn2]{present address: Dipartimento di Fisica, Università di Pisa, I-56127 Pisa, Italy 
and INFN Sezione di Pisa, I-56127 Pisa, Italy}

\begin{abstract}
	The response of a position-sensitive Li-glass scintillator detector to $\alpha$-particles
	from a collimated $^{241}$Am source scanned across the face of the detector has been
	measured.
	Scintillation light was read out by an 8~$\times$~8 pixel multi-anode photomultiplier and the signal amplitude
	for each pixel has been recorded for every position on a scan. The pixel signal is strongly 
	dependent on position and in general several pixels will register a signal (a hit) above a given 
	threshold. The effect of this threshold on hit multiplicity is studied, with 
	a view to optimize the single-hit efficiency of the detector.
\end{abstract}


\end{frontmatter}

\newpage
\section{Introduction}
\label{section:introduction}

The European Spallation Source (ESS)~\cite{ess} will soon commence operations 
as the most powerful neutron source in the world. Highly efficient, position sensing 
neutron detectors are crucial to the scientific mission of ESS. The worldwide
shortage of $^{3}$He~\cite{kouzes09,shea10,zeitelhack12} has resulted in 
considerable effort being undertaken to develop new neutron-detector technologies. 
One such effort is the development of \underline{S}olid-state \underline{N}eutron 
\underline{D}etectors SoNDe~\cite{sonde,jaksch17,jaksch18} for high-flux 
applications, motivated by the desire for two-dimensional position-sensitive 
systems for small-angle neutron-scattering 
experiments~\cite{heiderich91,kemmerling01,kemmerling04a,kemmerling04b,jaksch14,feoktystov15, ralf97, ralf98, ralf99, ralf02}. 
The SoNDe concept features individual neutron-detector modules which may be easily 
configured to instrument essentially any experimental phase space. The 
specification for the neutron interaction position reconstruction accuracy for the SoNDe 
technology is 6~mm. 

The core components of a SoNDe module are the 
neutron-sensitive Li-glass scintillator and the pixelated multi-anode 
photomultiplier tube (MAPMT) used to collect the scintillation light. The 
response of MAPMTs to scintillation-emulating laser light has been extensively 
studied~\cite{korpar00,rielage01,matsumoto04,lang05,abbon08,montgomery12,montgomery13,montgomery15,wang16}. 
Similar Li-glass/MAPMT detectors have been tested with thermal 
neutrons~\cite{zaiwei12} and a SoNDe detector prototype has been evaluated in a
reactor-based thermal-neutron beam~\cite{jaksch18}. In this paper, we present 
results obtained for the response of a SoNDe detector prototype to 
$\alpha$-particles from a collimated $^{241}$Am source, scanned across the face 
of the scintillator. Our goal was to examine methods to optimize the localization 
of the scintillation signal with a view to optimizing the position resolution of 
the detector, under constraints imposed by envisioned readout schemes for the 
detector. 
We were particularly interested in the behavior of the SoNDe detector
prototype at the vertical and horizontal boundaries between the pixels and the 
corners where four pixels meet.

\section{Apparatus}
\label{section:apparatus}

\subsection{Collimated $\alpha$-particle source}
\label{subsection:alpha_particle_beam}

\begin{figure}[H]
	\begin{center}
		\includegraphics[width=1.\textwidth]{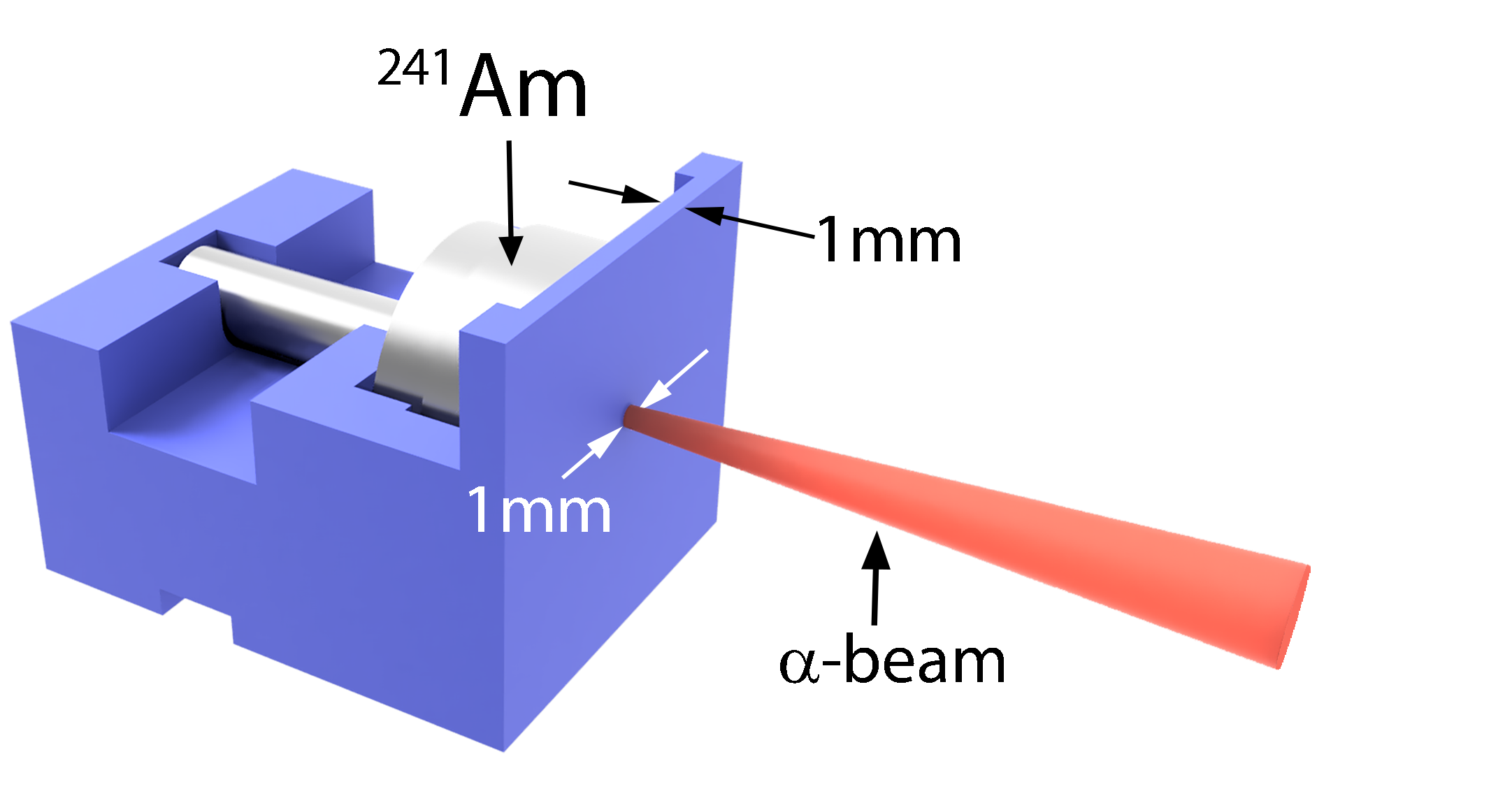}
		\caption{
			$^{241}$Am $\alpha$-particle source mounted in the 
			holder/collimator assembly used to define a beam of 
			$\alpha$-particles. The radioactive source is a snug fit in the blue holder/collimator. The assembly 
			shown has a 1~mm thick face plate and a 1~mm diameter 
			hole, resulting in the diverging red beam of 
			$\alpha$-particles. 
			(For interpretation of the references to color in this 
			figure caption, the reader is referred to the web version 
			of this article.)
			\label{figure:figure_01_HolderCollimator}
			}
	\end{center}
\end{figure}

Fig.~\ref{figure:figure_01_HolderCollimator} shows a sketch of the assembly 
used to produce a beam of $\alpha$-particles. It consisted of a
$^{241}$Am $\alpha$-particle source mounted in a 3D-printed holder/collimator 
assembly.

\begin{figure}[H]
	\begin{center}
		\includegraphics[width=1.\textwidth]{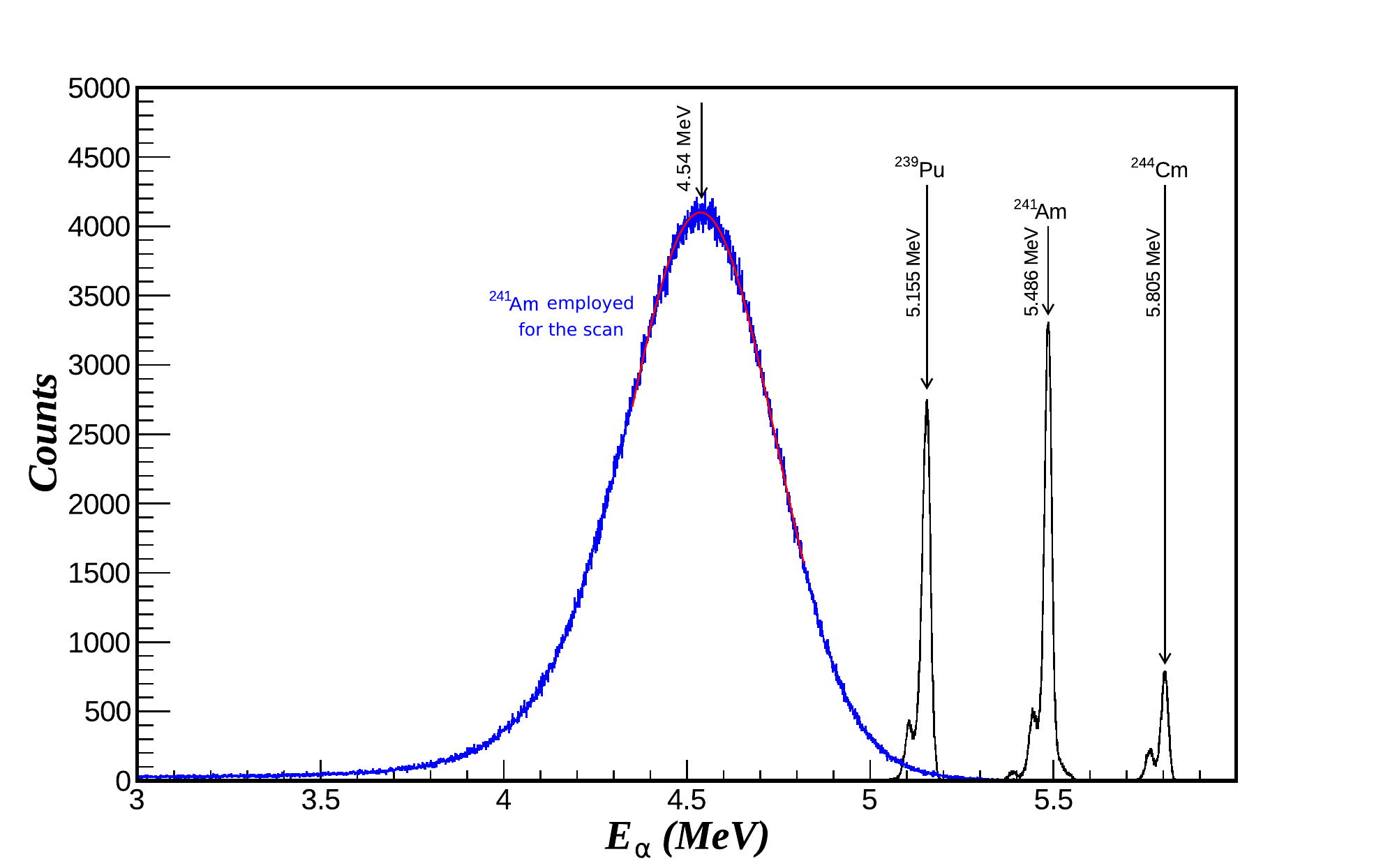}
		\caption{
			Uncollimated $\alpha$-particle spectrum emitted by the 
			$^{241}$Am source employed in the suite of measurements 
			reported on here (broad blue distribution) together with 
			a red fitted Gaussian function (indicating peaking at 
			4.54~MeV) and a triple $\alpha$-particle spectrum emitted 
			by a very thin-windowed three-actinide calibration source 
			(sharp black primary peaks at 5.155~MeV, 5.486~MeV, and 
			5.805~MeV).  
			(For interpretation of the references to color in this 
			figure caption, the reader is referred to the web version 
			of this article.)
			\label{figure:figure_02_AlphaCalib}
			}
	\end{center}
\end{figure}

The $\alpha$-particle emission energies of $^{241}$Am are 5.5~MeV ($\sim$85\%) 
and 5.4~MeV ($\sim$15\%). The gamma-ray background from the 
subsequent decay of excited states of $^{237}$Np has an energy of $\sim$60~keV and 
has a neglible effect on the $\alpha$-particle response of the SoNDe detector 
prototype. The $\alpha$-particle spectrum from the present source was measured 
(Fig.~\ref{figure:figure_02_AlphaCalib}) using a high-resolution passive-implanted 
planar silicon (PIPS) detector system in vacuum, where the PIPS detector was 
calibrated using a three-actinide calibration source. The average energy of the 
$\alpha$-particles emitted by the presently employed source was $\sim$4.5~MeV, 
corresponding to an energy loss of $\sim$1~MeV in the source window.

Holder/collimator assemblies for the $^{241}$Am source were 3D-printed from
polyactic acid using the fused deposition modeling technique. For our 
measurements, we used 1~mm thick collimators with either 3~mm (for gain mapping)
or 1~mm (for border scanning) diameter apertures, resulting in uniform 3~mm or 
1~mm irradiation spots at the upstream face of the scintillator. The distance from 
the $^{241}$Am source to this upstream face was $\sim$6~mm, so that the mean 
$\alpha$-particle energy at the surface of the GS20 wafer was 
$\sim$4.0~MeV~\cite{alpha_loss}.

\subsection{SoNDe detector prototype}
\label{subsection:detector}

The SoNDe detector prototype investigated in this paper is being developed for 
large-area arrays to detect thermal to cold neutrons with energies of $\leq$25~meV. 
It consists of a 1~mm thick lithium-silicate scintillating glass wafer coupled to an MAPMT.
\subsubsection{Li-glass scintillator}
\label{subsubsection:Liglass_scintillator}

Cerium-activated lithium-silicate glass scintillator 
GS20~\cite{firk61,spowart76,spowart77,fairly78} 
purchased from Scintacor~\cite{scintacor} was chosen for this application. 
GS20 has been demonstrated to be an excellent scintillator for 
the detection of thermal and cold neutrons and arrays of scintillator tiles can be arranged 
into large area detector systems~\cite{kemmerling04a,kemmerling04b,heiderich91,kemmerling01,feoktystov15}. 
The lithium content is 6.6\% by weight, with a 95\% $^{6}$Li isotopic enhancement, giving a $^{6}$Li 
concentration of 1.58~$\times$~$10^{22}$~atoms/cm$^{3}$. $^{6}$Li has a 
thermal-neutron capture cross section of $\sim$940~b at 25~meV, so that a 1~mm thick wafer of GS20 
detects $\sim$75\% of incident thermal neutrons. The process 
produces a 2.05~MeV $\alpha$-particle and a 2.73~MeV triton which have mean ranges of 
5.3~$\mu$m and 34.7~$\mu$m respectively~\cite{jamieson15} in GS20. Using the present 
$\alpha$-particle source, scintillation light was generated overwhelmingly within 
$\sim$15~$\mu$m of the upstream face of the scintillating wafer. We note that the 
scintillation light-yield outputs for 1~MeV protons is $\sim$5 times higher than that of 1~MeV $\alpha$-particles~\cite{dalton87}.
Thus, after thermal-neutron capture, the triton will produce a factor of 
$\sim$5 more scintillation light than the $\alpha$-particle.
Tests by van~Eijk~\cite{vanEijk04}
indicate $\sim$6600 scintillation photons per neutron event with a peak at a 
wavelength of 390~nm, which corresponds to $\sim$25\% of the anthracene benchmark. 
The sensitivity of GS20 to gamma-rays is energy dependent. A threshold cut will eliminate the low-energy gamma-rays, but higher-energy gamma-rays 
can produce large pulses if the subsequent electrons (from Compton scattering or pair production) traverse sufficient thickness of GS20.

Our glass wafer was 1~mm thick and 50~mm~$\times$~50~mm in 
area. The glass faces, apart from the edges, were polished and the wafer was fitted to the MAPMT window without any optical coupling medium. The index of refraction of GS20 is 1.55 at 395~nm. We have 
assumed that the $^6$Li distribution in our scintillating wafer was uniform.

\subsubsection{Multi-anode photomultiplier tube}
\label{subsubsection:mapmt}
\begin{figure}[H]
	\begin{center}
		 \begin{subfigure}[b]{0.57\textwidth}
                        \includegraphics[width=1.\textwidth]{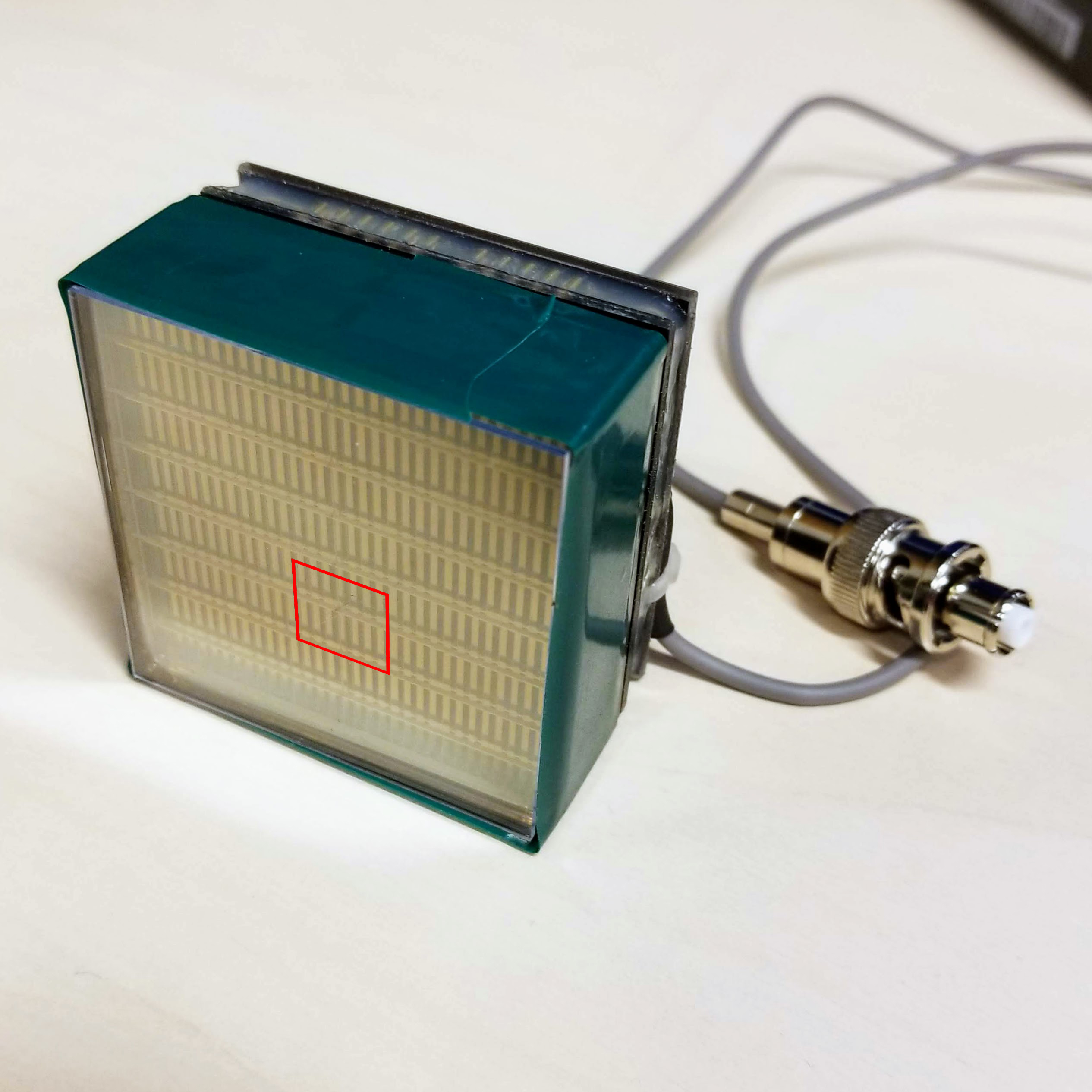}
                        \caption{MAPMT with sctintillator}
                        \label{subfig:photograph}
                \end{subfigure}
		 \begin{subfigure}[b]{0.57\textwidth}
                        \includegraphics[width=1.\textwidth]{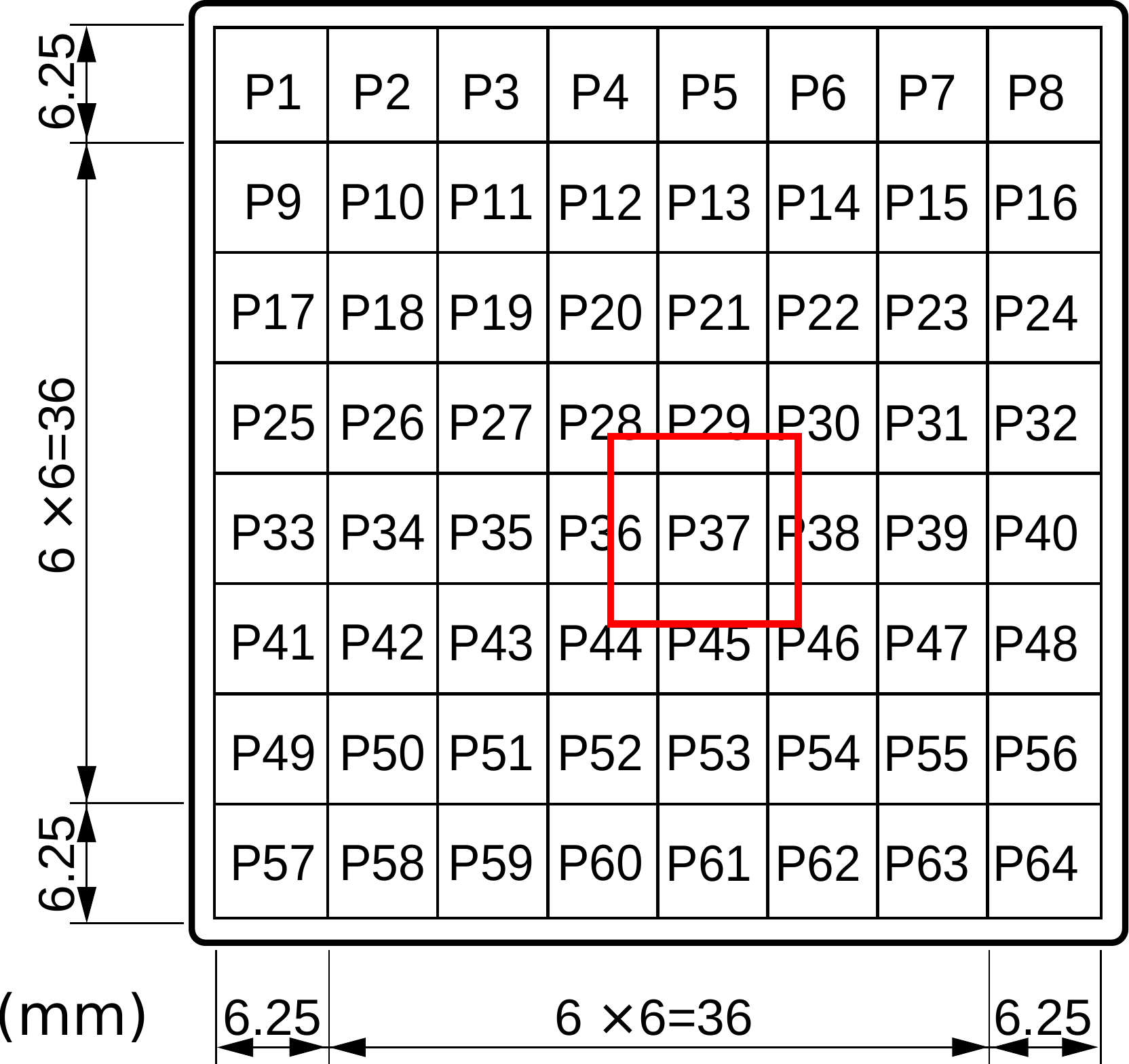}
                        \caption{MAPMT pixel map}
                        \label{subfig:pixelmap}
                \end{subfigure}
		\caption{
			The Hamamatsu 12700 MAPMT. \ref{subfig:photograph}: 
			Photograph of the MAPMT together with the GS20 
			scintillator wafer. \ref{subfig:pixelmap}: 
			Numbering of the 64 MAPMT pixels (front view). Pixel~1 
			(P1) is located in the top left-hand corner of the 
			MAPMT looking into it from the front. Sketch from Ref.~\cite{hamabrochure}. The red 
			boxes indicate the region of irradiation reported on in 
			detail in this paper.
			(For interpretation of the references to color in this 
			figure caption, the reader is referred to the web version 
			of this article.)
			\label{figure:figure_XX_MAPMTDetector}}
	\end{center}
\end{figure}

The Hamamatsu type H12700 MAPMT employed in the SoNDe detector prototype is an 
8~$\times$~8 pixel device with outer dimensions 52~mm~$\times$~52~mm and an active 
cathode area of 48.5~mm~$\times$~48.5~mm, resulting in a packing density of 87\%. 
The bialkali photocathode produces a peak quantum efficiency of $\sim$33\% at
$\sim$380~nm wavelength, which is well matched to the GS20 scintillation.
Compared to its predecessor type H8500 MAPMT, the H12700 MAPMT achieves similar overall
gain, but with 10 as opposed to 12 dynode stages. The H12700 MAPMT employed for the
present tests had a gain of 2.09~$\times$~10$^{6}$ and a dark current of 2.67~nA
at an anode-cathode potential of $-$1000~V.
Each of the 64 $\sim$6~$\times$~6~mm$^2$ pixels in the Hamamatsu 12700 MAPMT has a slightly different gain,
which is measured and documented by the supplier. A typical H12700 MAPMT has a factor 
2 variation in pixel gain (factor 3 worst case)~\cite{hamabrochure}. The 
datasheet provided by Hamamatsu for the H12700 MAPMT used in this study had a
worst case anode-to-anode gain difference of a factor 1.7. 
Figure~\ref{figure:figure_XX_MAPMTDetector} shows a photograph of the device 
together with a pixel map.

\section{Measurement}
\label{section:measurement}

The SoNDe detector prototype was irradiated using the collimated beams of 
$\alpha$-particles (Sec.~\ref{subsection:alpha_particle_beam}) where the center of 
the beam was directed perpendicular to the face of the GS20 wafer. The downstream 
face of the source holder/collimator assembly was translated parallel to the 
surface of the scintillator wafer on an XY-coordinate scanner, powered by a pair 
of Thorlabs NRT150 stepping motors~\cite{thorlabs}. This was programmed to scan a 
lattice of irradiation points uniformly distributed across the face of the device. 
The entire assembly was located within a light-tight box and the temperature 
($\sim$25\degree), pressure ($\sim$101.3~kPa), and humidity ($\sim$30\%) within 
the box were logged at the beginning and end of each measurement position. The 
anode signals from each of the pixels in the MAPMT were recorded using standard 
NIM and VME electronics. The positive polarity dynode-10 signal was shaped and 
inverted using an ORTEC~454 NIM timing filter amplifier producing a negative 
polarity signal with risetime $\sim$5~ns, falltime $\sim$20~ns, and amplitude some 
tens of mV, which was fed to an ORTEC CF8000 constant-fraction discriminator set 
to a threshold of $-$8~mV. The resulting NIM logic pulses with a duration of 150~ns 
provided a trigger for the data-acquisition system and a gate for the CAEN V792 VME 
charge-to-digital converters (QDCs) used to record the 64 anode pixel charges. A 
CAEN V2718 VME-to-PCI optical link bridge was used to sense the presence of a 
trigger signal and to connect the VMEbus to a Linux PC-based data-acquisition 
system. The digitized signals were recorded on disk and subsequently processed using 
ROOT-based software~\cite{root}. Data were recorded for $\sim$120~s at each point on 
a scan, so that in total a scan could take several hours.

\section{Results}
\label{section:results} 
\begin{figure}[H]
	\begin{center}
                \begin{subfigure}[b]{0.73\textwidth}
                        \includegraphics[width=1.\textwidth]{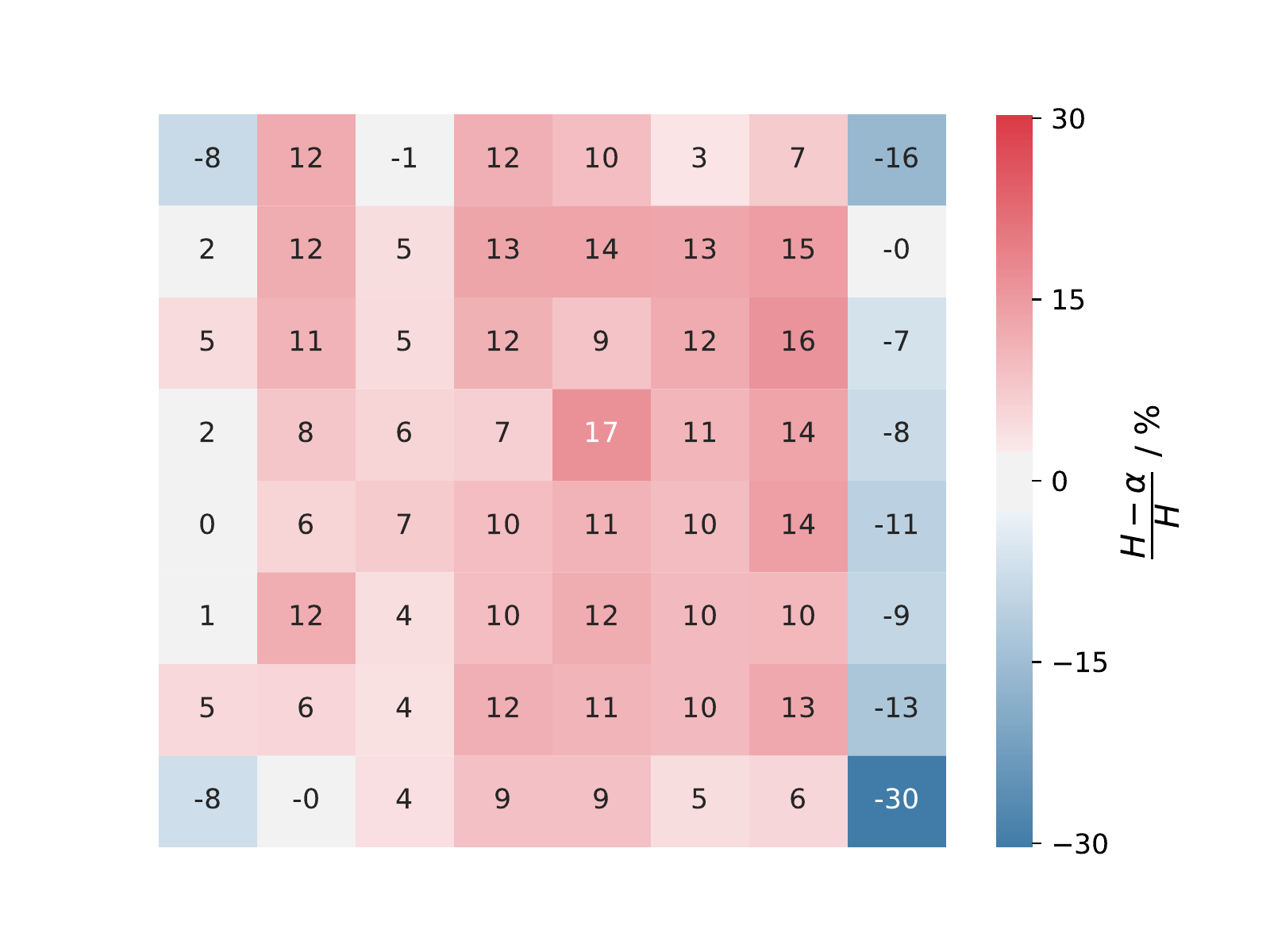}
                        \caption{Gain differences, areal}
                        \label{subfig:heatmap}
                \end{subfigure}
                \begin{subfigure}[b]{0.73\textwidth}
                        \includegraphics[width=1.\textwidth]{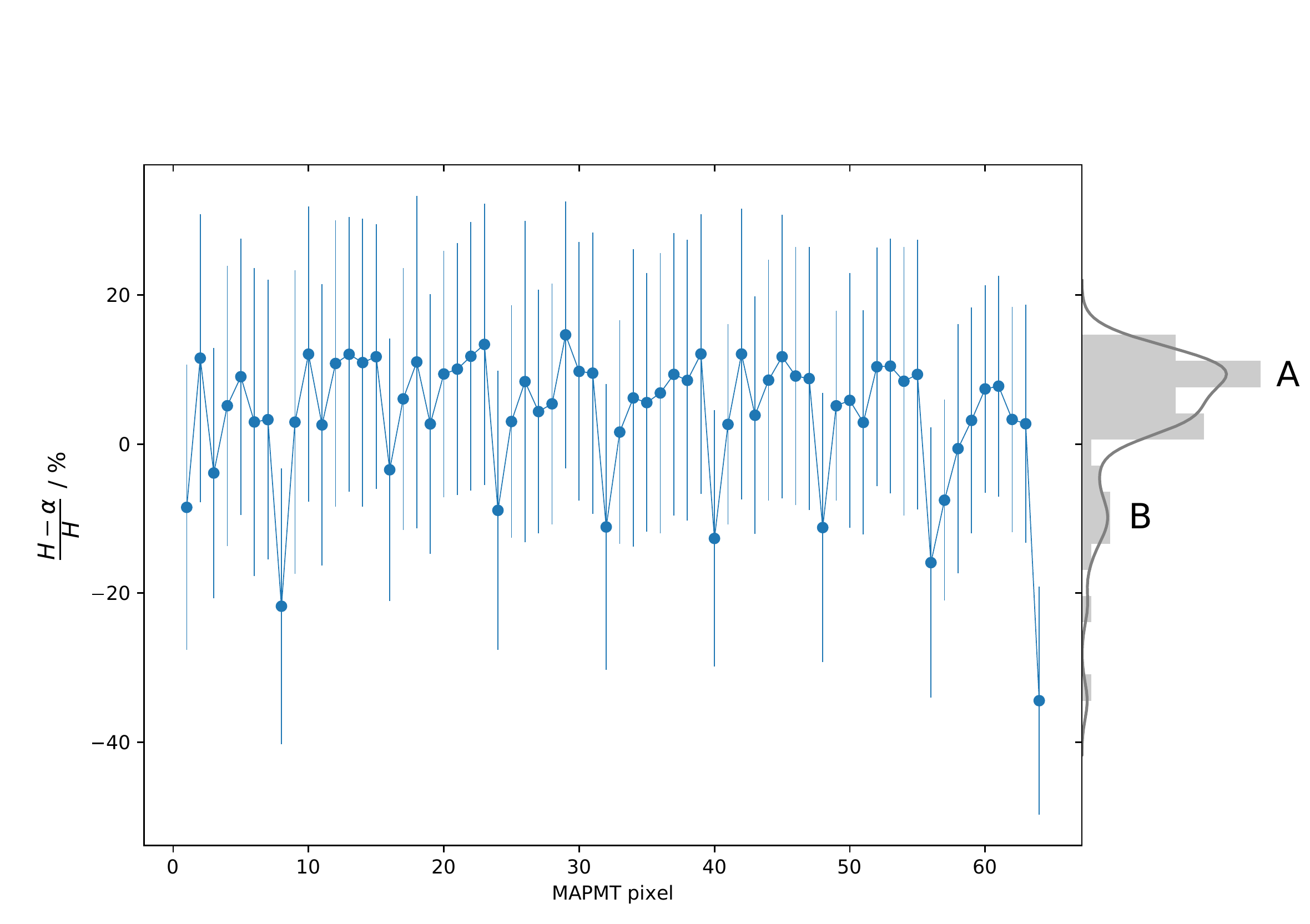}
                        \caption{Gain differences, projected}
                        \label{subfig:heatmap_projection}
                \end{subfigure}
		\caption{
			Differences between the $\alpha$-scan gain-map and the 
			Hamamatsu gain-map normalized to the Hamamatsu gain-map 
			in percent.
			\ref{subfig:heatmap}: 2D representation in which 
			the top-left corner 
			corresponds to P1. \ref{subfig:heatmap_projection}: 
			1D representation of the 
			same as a function of pixel. Error bars are derived from
			fit widths. The values have been joined with a line to 
			guide the eye. A histogram of the gain differences is 
			projected in grey on the right vertical axis. Cluster~A of that histogram
			corresponds to red pixels in \ref{subfig:heatmap} 
			while cluster~B corresponds to blue pixels.  
			(For interpretation of the references to color in this 
			figure caption, the reader is referred to the web version 
			of this article.)
			}
		\label{figure:gainmaps}
	\end{center}
\end{figure}

The gain-calibration datasheet provided by Hamamatsu may be used to correct for 
non-uniform response. However, previous 
work~\cite{korpar00,rielage01,matsumoto04,lang05,abbon08,montgomery12,montgomery15} 
has clearly suggested that mapping of pixel gains is highly dependent upon the irradiation 
conditions. Since our $\alpha$-particle beam results in very short ($\sim$10s~of~ns) 
pulses of highly localized scintillation light, which are in sharp contrast to the 
steady-state irradiation measurement employed by Hamamatsu, we re-measured 
the gain-map of our MAPMT {\it in situ} using the equipment described previously. 
For each pixel, the 3~mm diameter $\alpha$-particle collimator was centered on the 
XY position of the pixel center of the MAPMT photocathode and 2200 $\alpha$-particle 
events were recorded. The resulting anode-charge distributions were well fitted with
Gaussian functions (average $\chi^2$ per degree 
of freedom~1.50 with a variance of 0.14) and the means, $\mu$, and 
standard deviations, $\sigma$, were recorded. The largest measured $\mu$-value 
(corresponding to the pixel with the highest gain) was normalized to 100.
The relative difference between the Hamamatsu gain-map values, $H$, and the $\alpha$-scanned 
gain-map values, $\alpha$, was calculated as $\frac{H - \alpha}{H}$ on a pixel-by-pixel basis.
Figure~\ref{figure:gainmaps} 
shows the results of our gain-map measurement, where the present results show significant differences to the Hamamatsu measurement. 
General trends in regions of high and low gain agree.
Measurements of a sample of 30 H12700 MAPMTs revealed that the
window face is not flat and was systematically $\sim$80~$\mu$m lower at the center
compared to the edges. The non-uniformity of the air gap between the GS20 and MAPMT
window may be a partial cause of the gain discrepancy displayed in 
Fig.~\ref{figure:gainmaps}, as may reflections at the edges of the GS20 wafer.
In the following, we use the present $\alpha$-scintillation generated gain-map, 
which in principle will embody non-uniform light-collection effects. Nonetheless,
scintillation-light propagation through the SoNDe detector protoype is being 
studied in a \geant-based simulation~\cite{agostinelli03,allison06} in order to 
better understand these effects.

\begin{figure}[H]
        \begin{center}
        	\begin{subfigure}[b]{0.32\textwidth}
                	\includegraphics[width=0.99\textwidth]{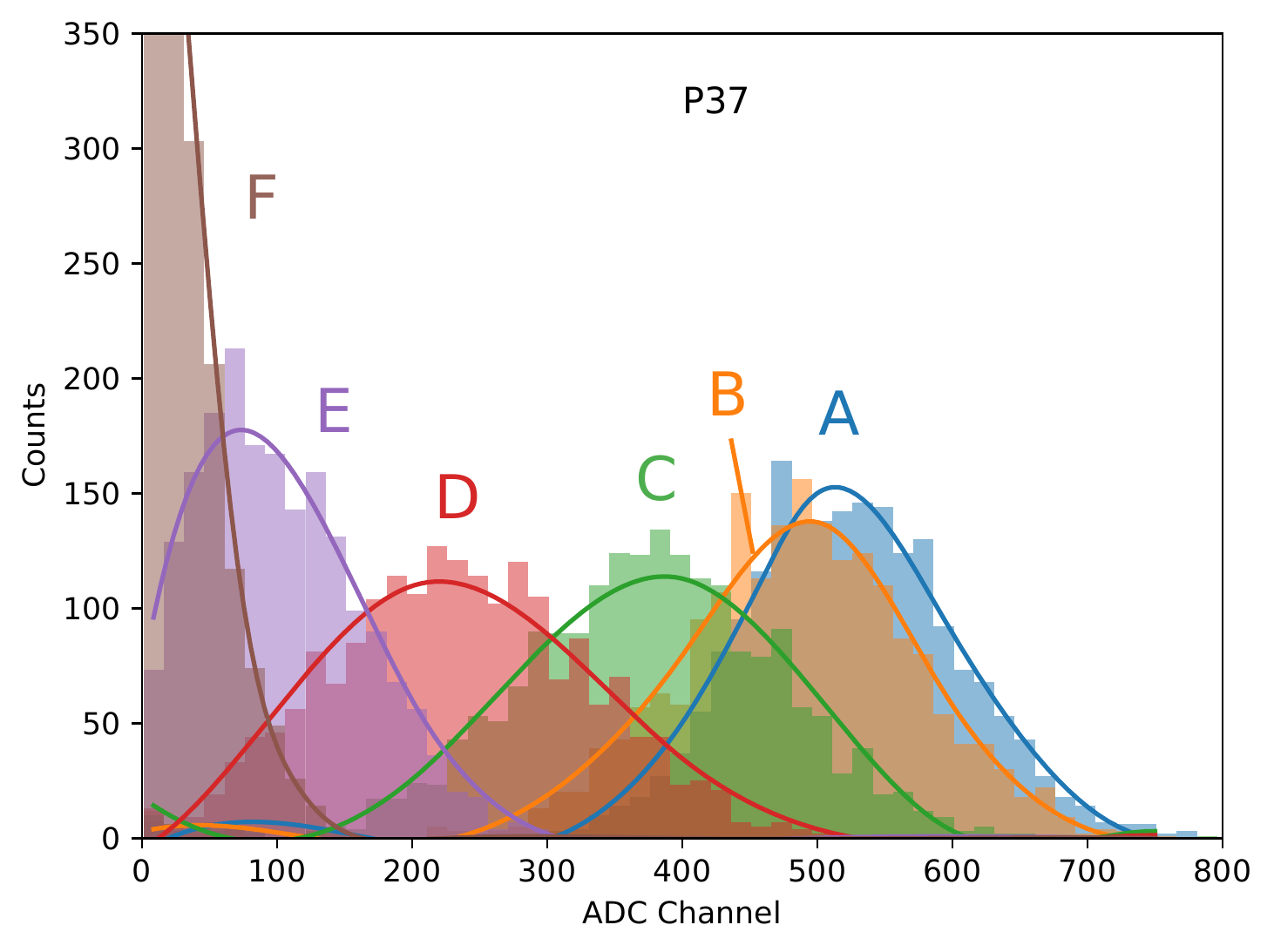}
        	\end{subfigure}
        	\hfill
        	\begin{subfigure}[b]{0.32\textwidth}
                	\includegraphics[width=0.99\textwidth]{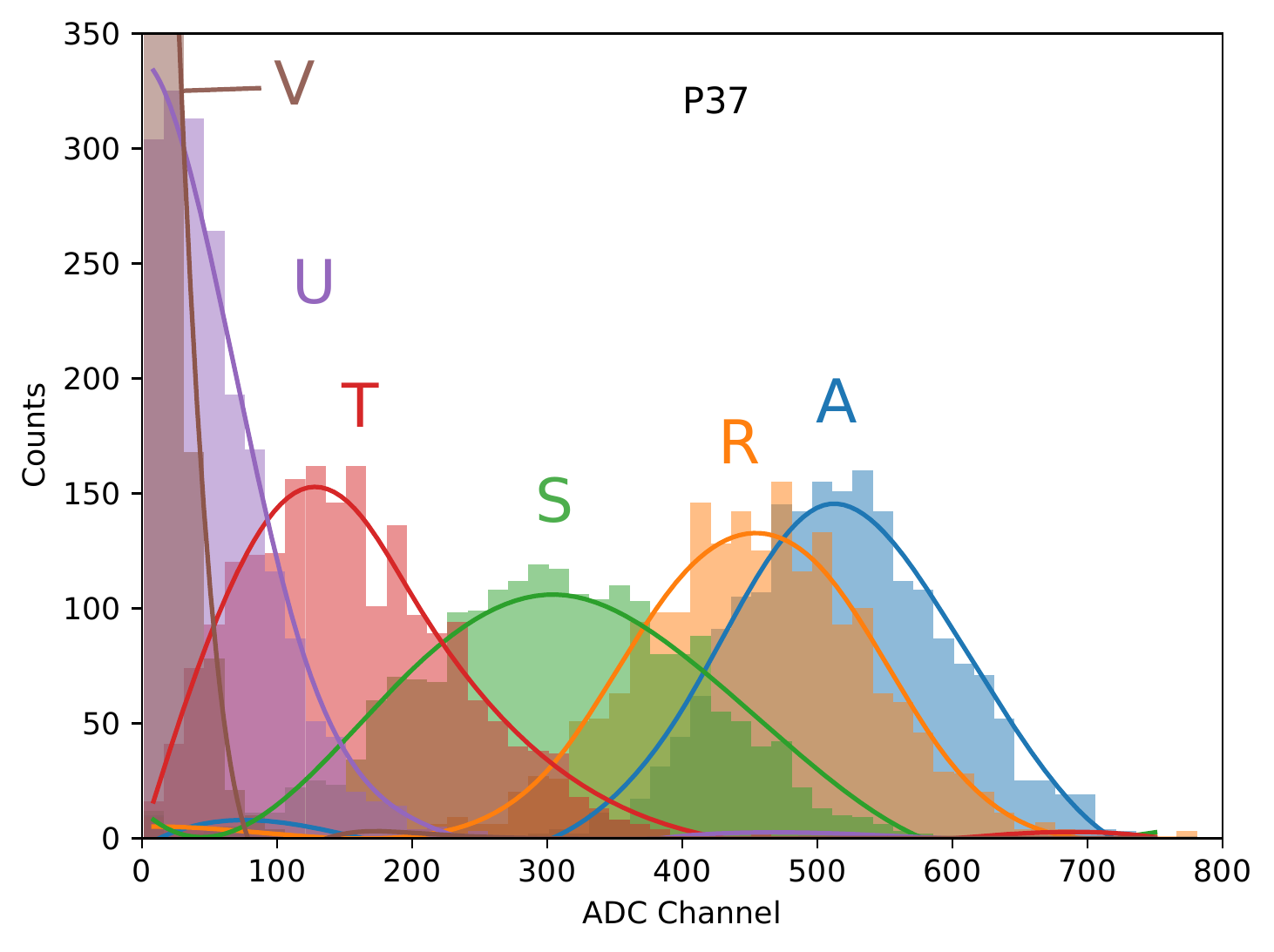}
        	\end{subfigure}
        	\hfill
        	\begin{subfigure}[b]{0.32\textwidth}
                	\includegraphics[width=0.99\textwidth]{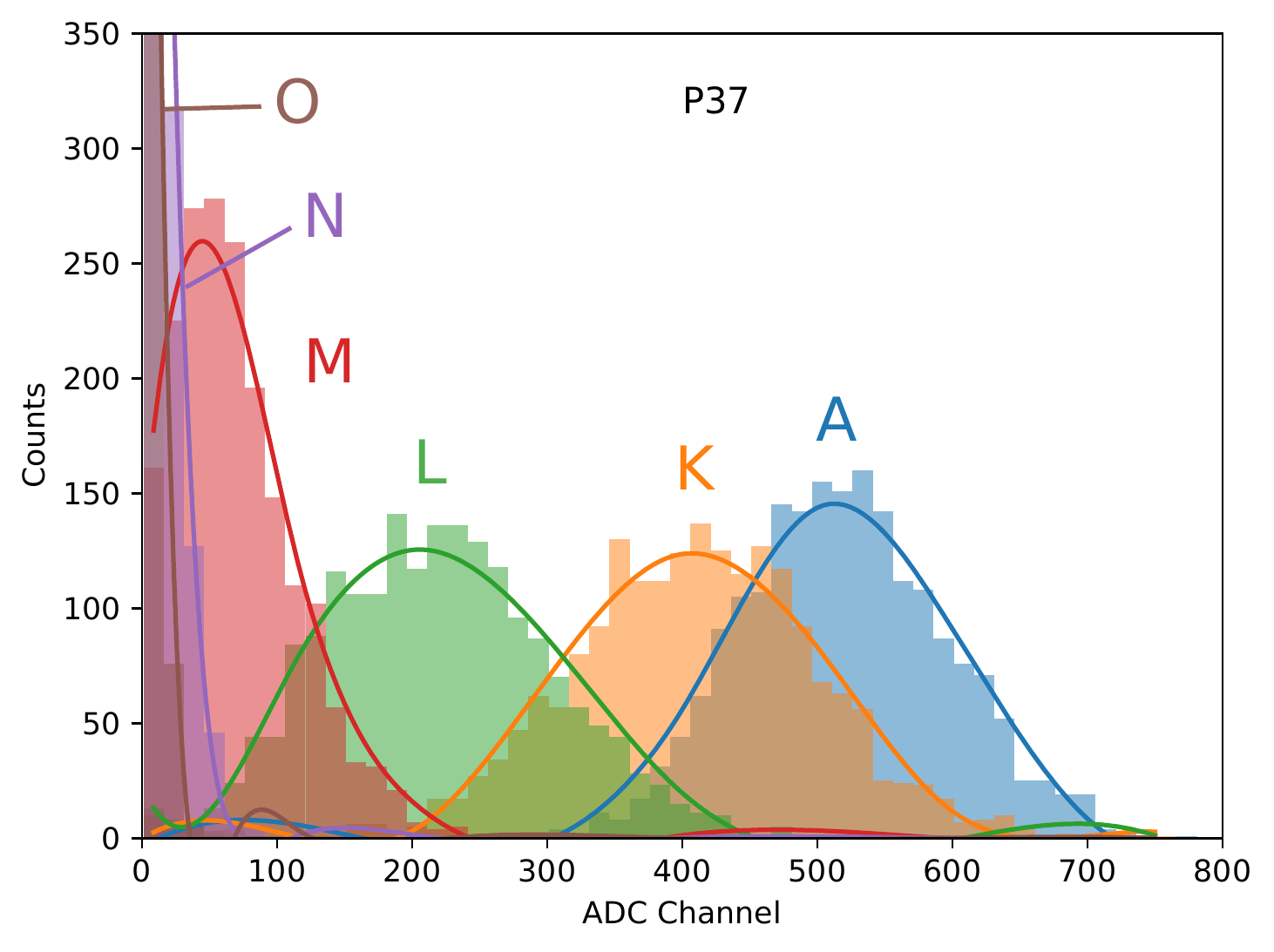}
        	\end{subfigure}
        	\quad
        	\begin{subfigure}[b]{0.32\textwidth}
                	\includegraphics[width=0.99\textwidth]{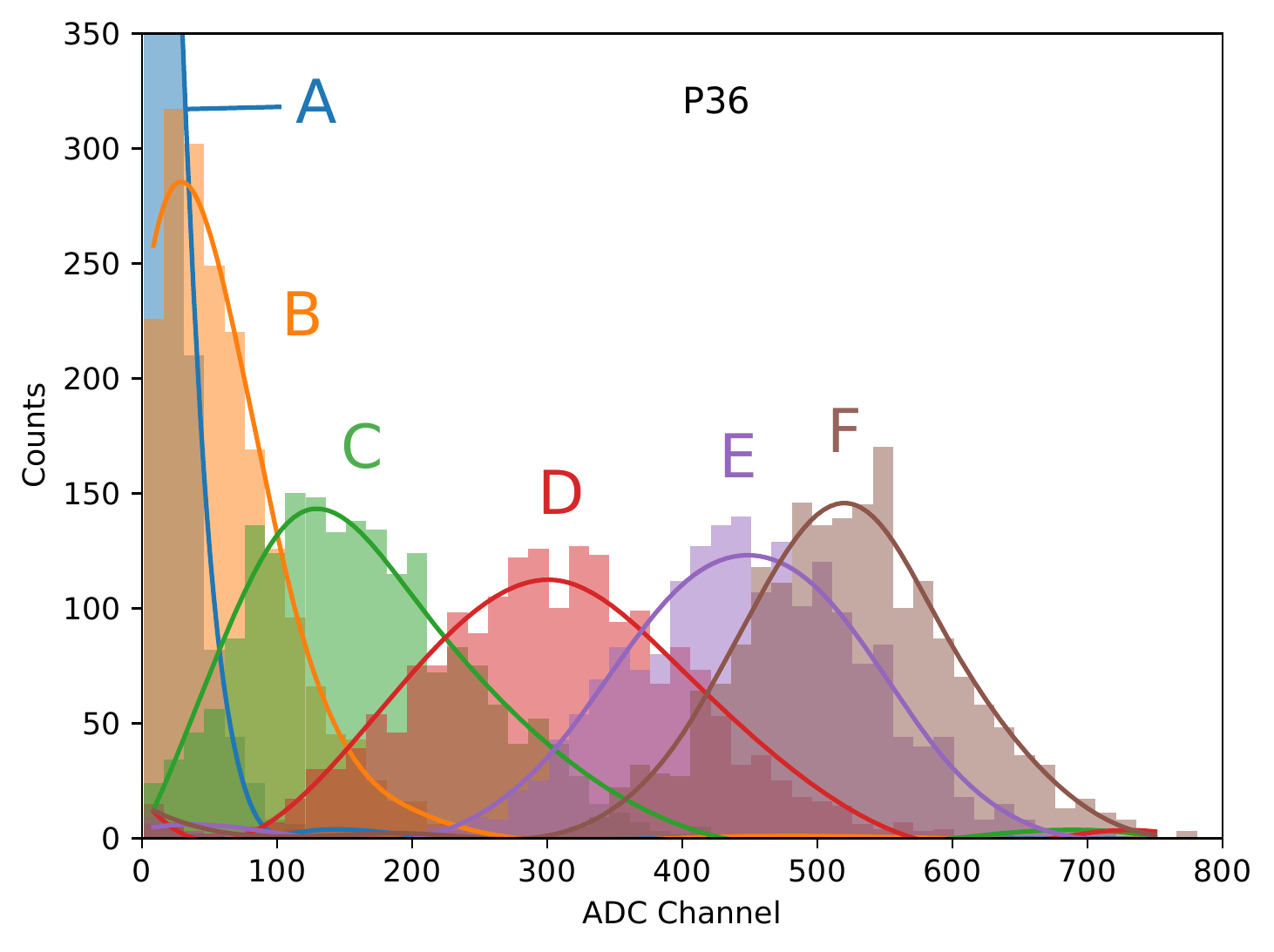}
			\caption{Horizontal scan F $\leftarrow$ A}
                	\label{subfig:horscan}
        	\end{subfigure}
        	\hfill
        	\begin{subfigure}[b]{0.32\textwidth}
                	\includegraphics[width=0.99\textwidth]{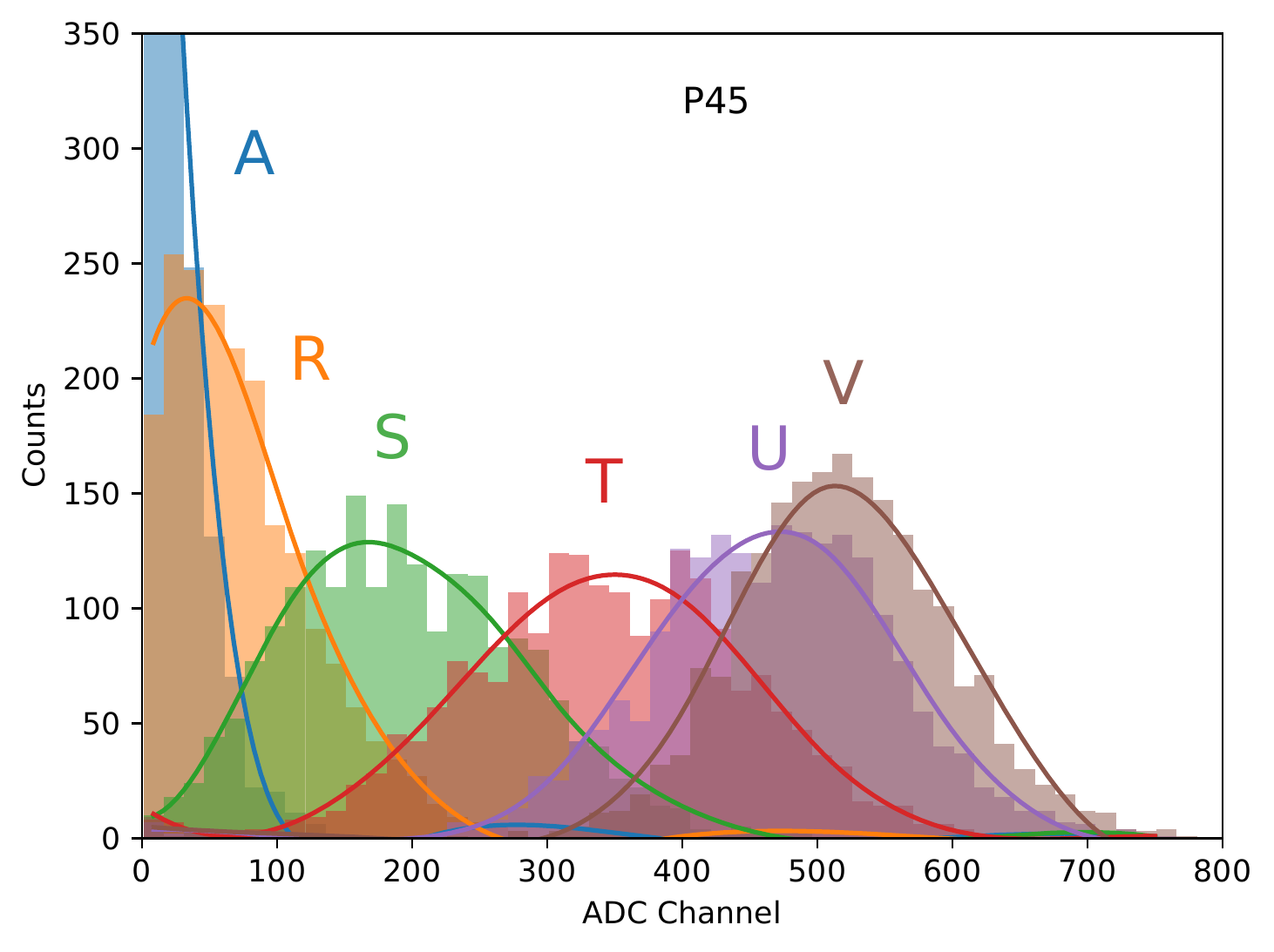}
			\caption{Vertical scan A $\downarrow$ V}
                	\label{subfig:verscan}
        	\end{subfigure}
        	\hfill
        	\begin{subfigure}[b]{0.32\textwidth}
                	\includegraphics[width=0.99\textwidth]{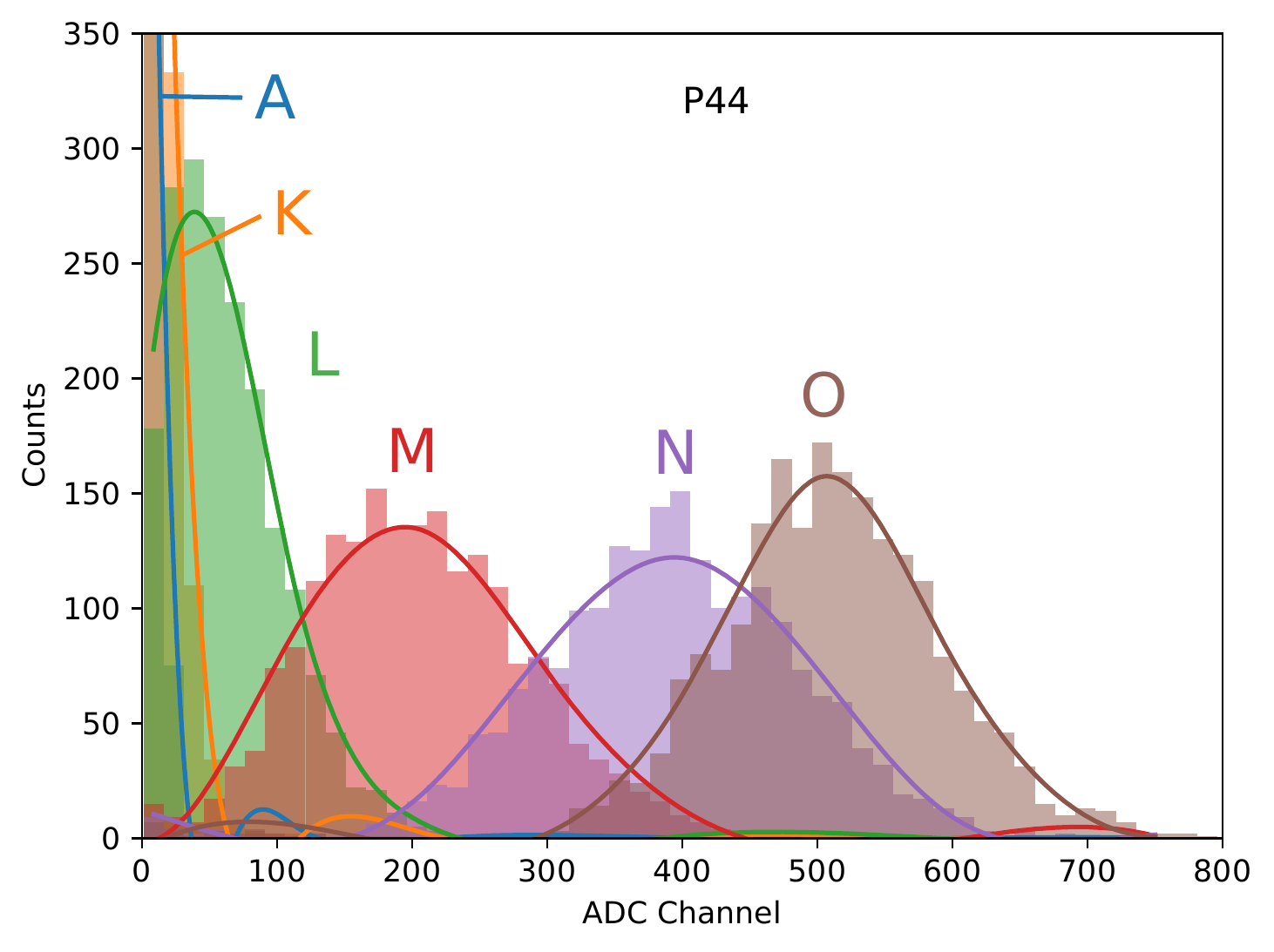}
			\caption{Diagonal scan O $\swarrow$ A}
                	\label{subfig:diascan}
        	\end{subfigure}
		\begin{subfigure}[b]{0.43\textwidth}
                	\hspace*{5mm}
                	\includegraphics[width=0.75\textwidth]{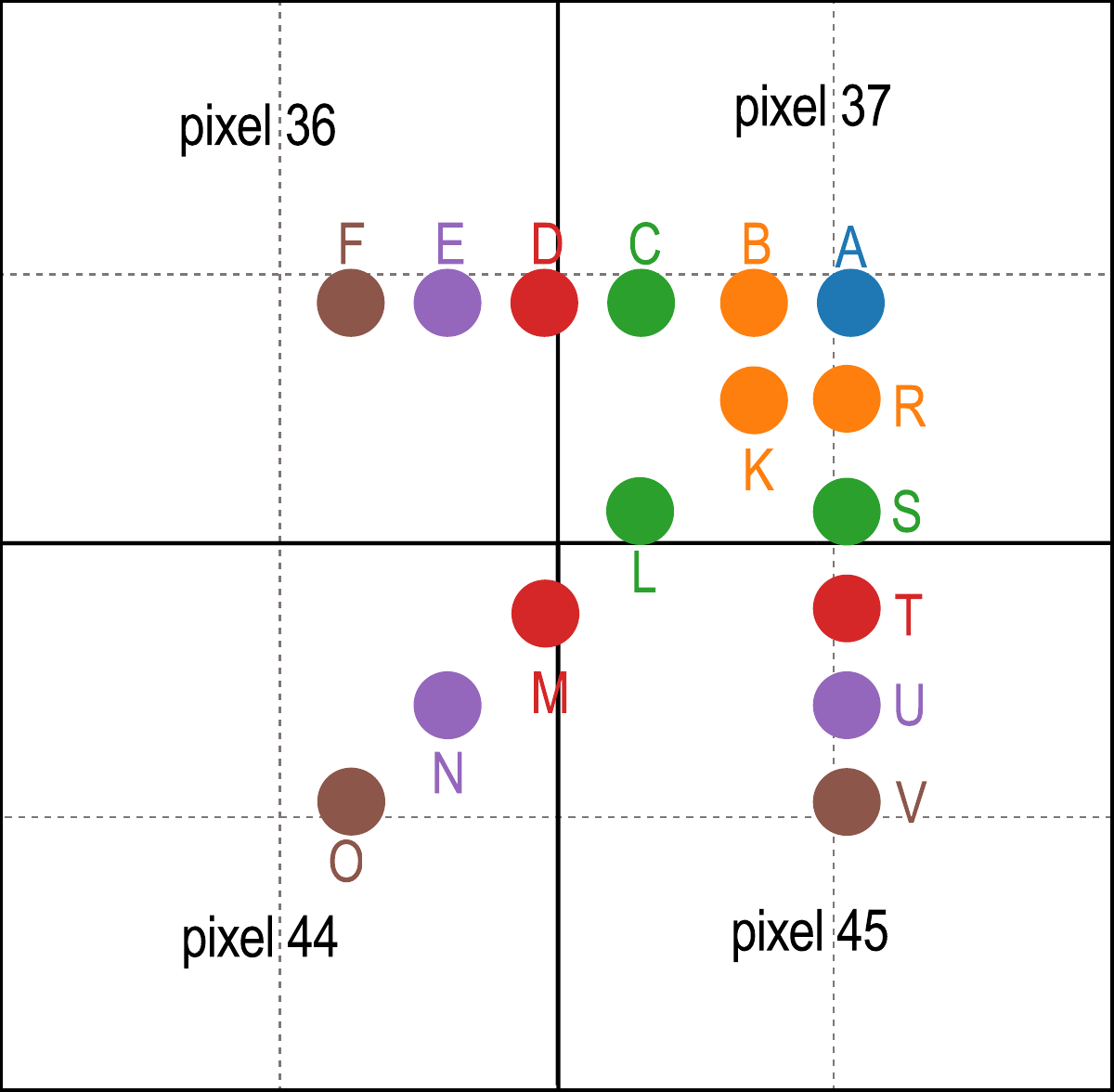}
               		\vspace*{6mm}
			\caption{Key}
                        \label{subfig:key}
                \end{subfigure}
		\begin{subfigure}[b]{0.56\textwidth}
                	\hspace*{-5mm}
                        \includegraphics[width=1.0\textwidth]{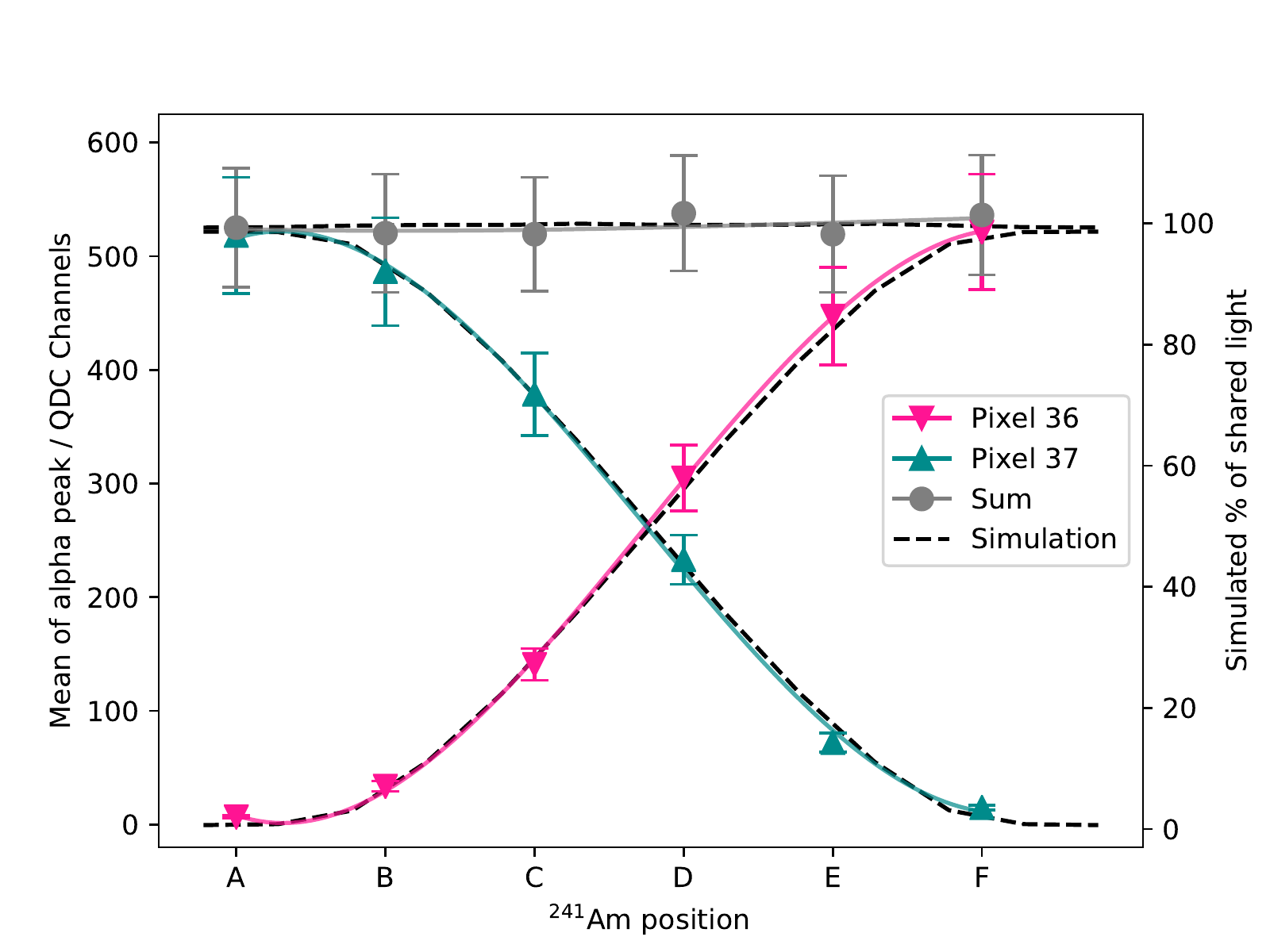}
			\caption{Light sharing horizontal scan F $\leftarrow$ A}
                        \label{subfig:lightshare}
                \end{subfigure}
		\caption{
			\ref{subfig:horscan}--\ref{subfig:diascan}: 
			Measured charge distributions for four adjacent pixels 
			as the $\alpha$-particle beam was translated in 1~mm 
			horizontal and vertical steps across the pixel 
			boundaries. (a): horizontal scan from P37 to 
			P36. (b): vertical scan from P37 to P45. 
			(c): diagonal scan from P37 to P44. 
			\ref{subfig:key}: Key. The solid (dashed) black 
			lines indicate the pixel boundaries (centers). 
			\ref{subfig:lightshare}: Gain-corrected means of 
			the collected charge distributions corresponding to 
			\ref{subfig:horscan}. The curves are splines 
			drawn to guide the eye. Error bars correspond to 
			$\sigma/10$. 
			(For interpretation of the references to color in 
			this figure caption, the reader is referred to the 
			web version of this article.)
			\label{figure:mean_yields_with_gain_correction}
			}
	\end{center}
\end{figure}

Figure ~\ref{figure:mean_yields_with_gain_correction} shows the positions of 
the 1~mm collimated $\alpha$-particle source employed for the horizontal, 
vertical, and diagonal XY scans. The positions are labeled A--V and color 
coded (\ref{subfig:key}). The $\alpha$-particle pulse-height spectra 
recorded at each of the scan positions are displayed in 
\ref{subfig:horscan}--\ref{subfig:diascan} for pixels P36, P37, P44, and P45 
which encompass the scan coordinates. The QDC pulse-height distributions 
have been pedestal subtracted and corrected for non-uniform pixel gain 
(Fig.~\ref{figure:gainmaps}). It is obvious that the efficiency of scintillation 
light collection in a single pixel is strongly dependent on the position of the 
$\alpha$-particle interaction. The signal amplitude is maximized when light is 
produced at the center of the pixel. This variation in amplitude with position 
may be seen more clearly in \ref{subfig:lightshare}, which shows the mean of 
the pulse-height distributions as a function of interaction position for the
horizontal scan (\ref{subfig:horscan}). The full curves are splines to guide the eye, while the dashed curves display the predictions of a ray-tracing
simulation of light propagation~\cite{ROS}. The simulation is in good 
agreement with the measured data. After gain correction, these distributions
should be symmetric about the pixel boundary locations. 
The system was aligned such that position D should have corresponded to the
boundary between P36 and P37. However, the fits to the data suggest that the 
scan positions were offset by 0.2~mm to the left (Fig.~\ref{subfig:lightshare}). 
Corresponding fits to the vertical-scan data show a 0.4~mm vertical offset. 
The sum of the means of the two adjacent pixels scanned is also displayed. This 
shows that the amount of light collected by the two pixels over which the scan 
is performed is independent of position.

In general, several pixels adjacent to the target pixel will collect some 
scintillation light and in principle, this could be used to better localize the
position of the scintillation as in an Anger camera~\cite{anger58,riedel15}. While possible, 
this will not be the standard mode of operation for SoNDe modules running at ESS due to data-volume 
limitations. For production running at ESS, MAPMT 
pixels will be read and time-stamped on an event-by-event basis as lying either 
above or below per-pixel discriminator thresholds.

\begin{figure}[H]
	\centering 
	\begin{subfigure}[b]{0.49\textwidth}
		\hspace*{10mm}\includegraphics[width=0.7\textwidth]{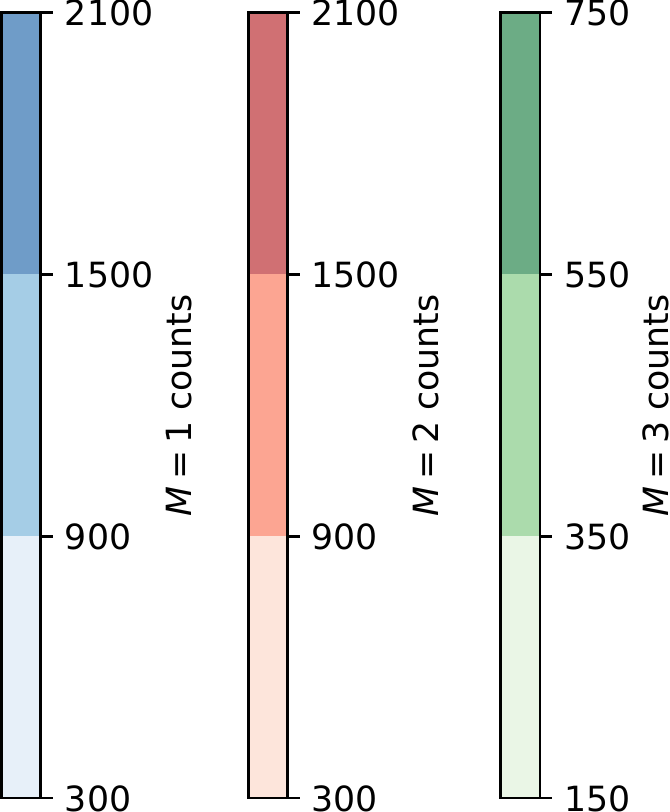}
		\vspace*{4mm}
		\caption{Key}
                \label{subfig:multkey}
	\end{subfigure}
	\begin{subfigure}[b]{0.49\textwidth}
		\includegraphics[width=1\textwidth]{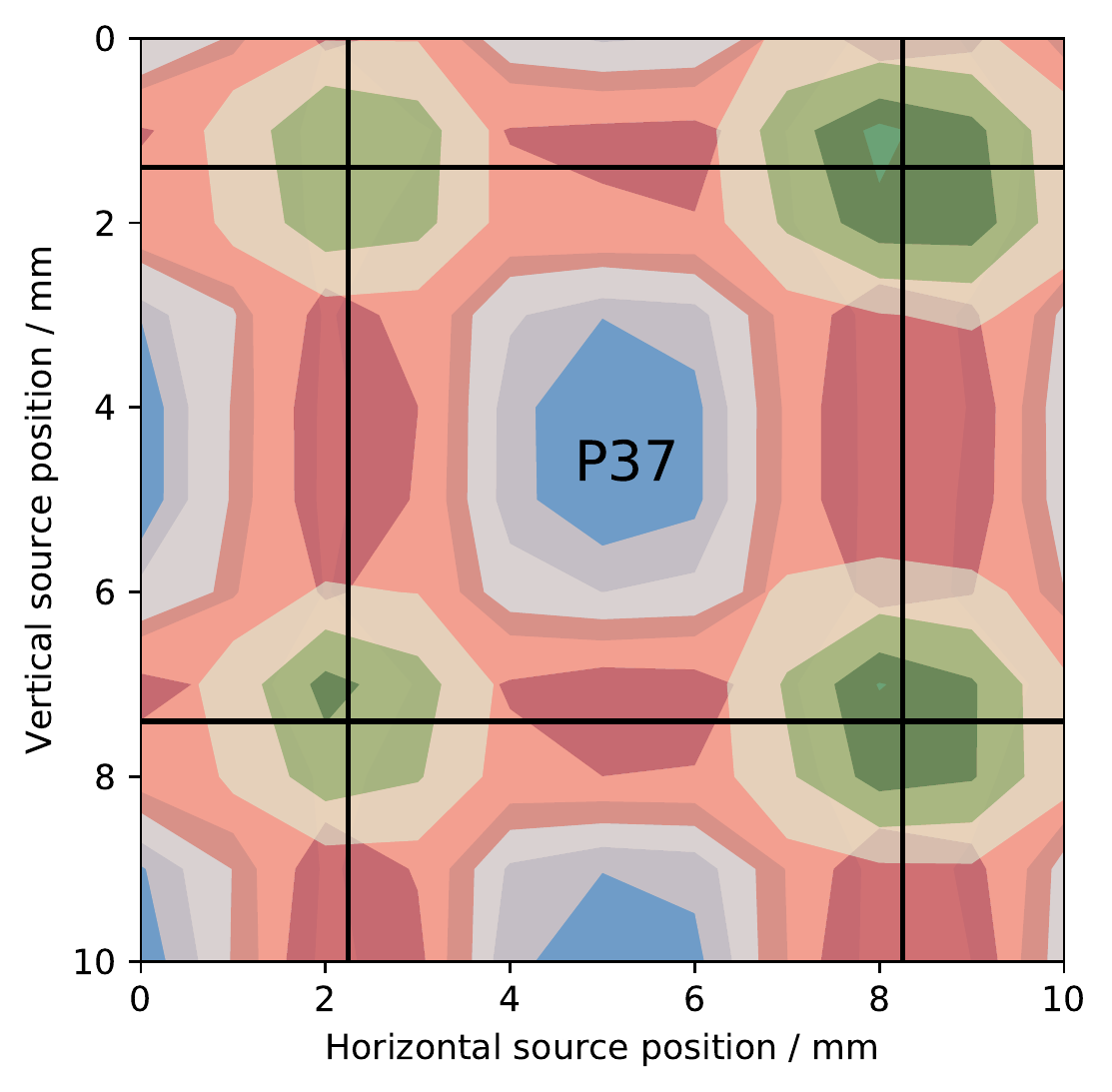}
		\vspace*{-8mm}
		\caption{Threshold 100 QDC~channels}
                \label{subfig:100_channels}
	\end{subfigure}
	\begin{subfigure}[b]{0.49\textwidth}
		\includegraphics[width=1\textwidth]{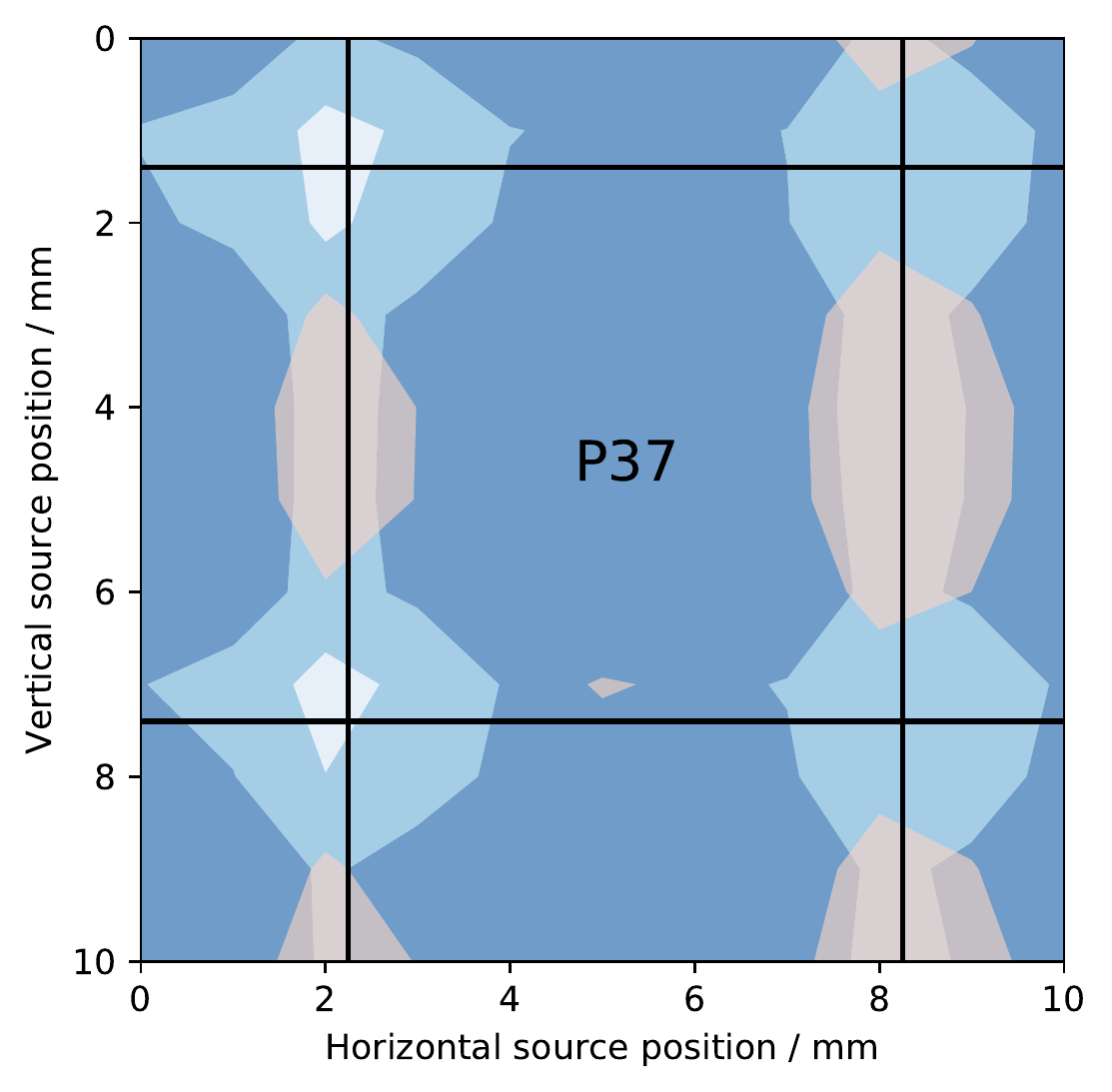}
		\vspace*{-8mm}
		\caption{Threshold 235 QDC~channels}
                \label{subfig:235_channels}
	\end{subfigure}
	\begin{subfigure}[b]{0.49\textwidth}
		\includegraphics[width=1\textwidth]{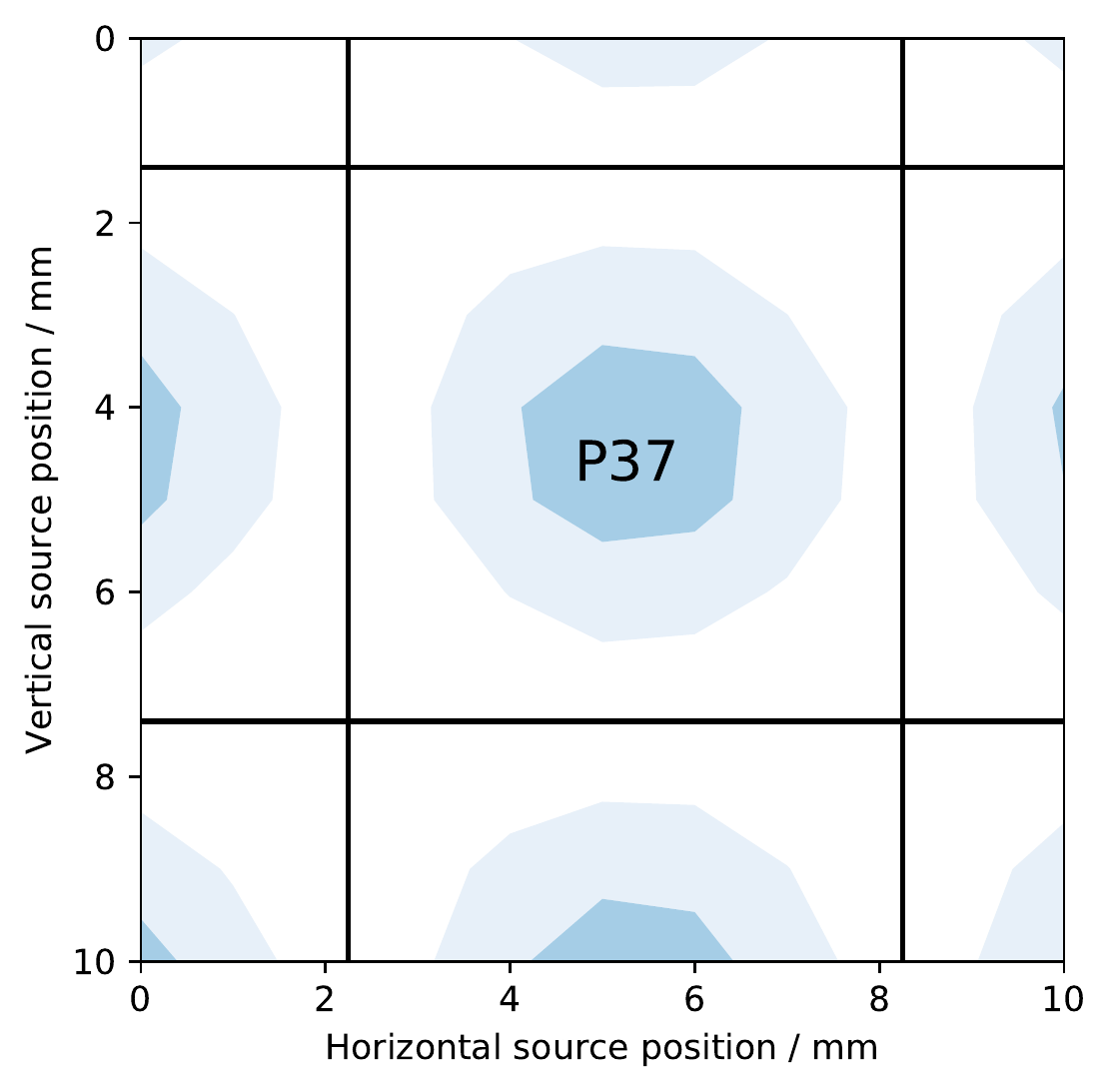}
		\vspace*{-8mm}
		\caption{Threshold 500 QDC~channels}
                \label{subfig:500_channels}
	\end{subfigure}
	\caption{
                Contour plots of the multiplicity distributions for pixels 
		lying near P37 for different thresholds as a function of 
		$\alpha$-particle beam irradiation location. The black lines 
		denote the pixel boundaries. In all three contour plots, blues 
		indicate $M$~$=$~1 events, reds indicate $M$~$=$~2 events, and 
		greens indicate $M$~$=$~3 events. The lighter the shade of the
		color, the fewer the counts.
		(For interpretation of the references to color in this
                figure caption, the reader is referred to the web version
                of this article.)
                \label{figure:contour_maps_M_with_cuts}
                }
\end{figure}

\begin{figure}[H]
        \begin{center}
	        \begin{subfigure}[b]{0.73\textwidth} 
			\includegraphics[width=1.\textwidth]{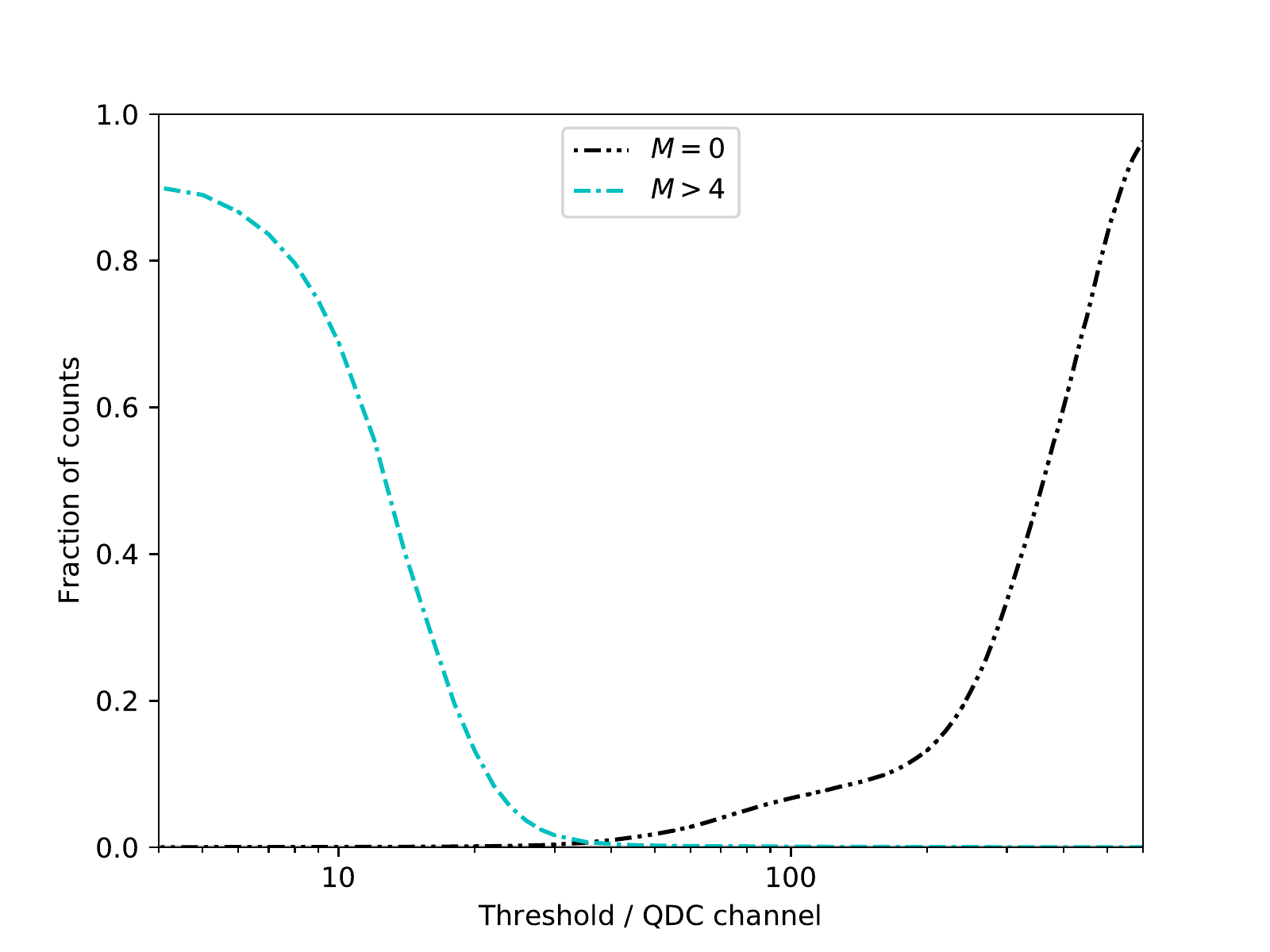}
			\caption{Extreme multiplicities}
                \label{subfig:m_extreme}
        	\end{subfigure}
                \begin{subfigure}[b]{0.73\textwidth}
                        \includegraphics[width=1.\textwidth]{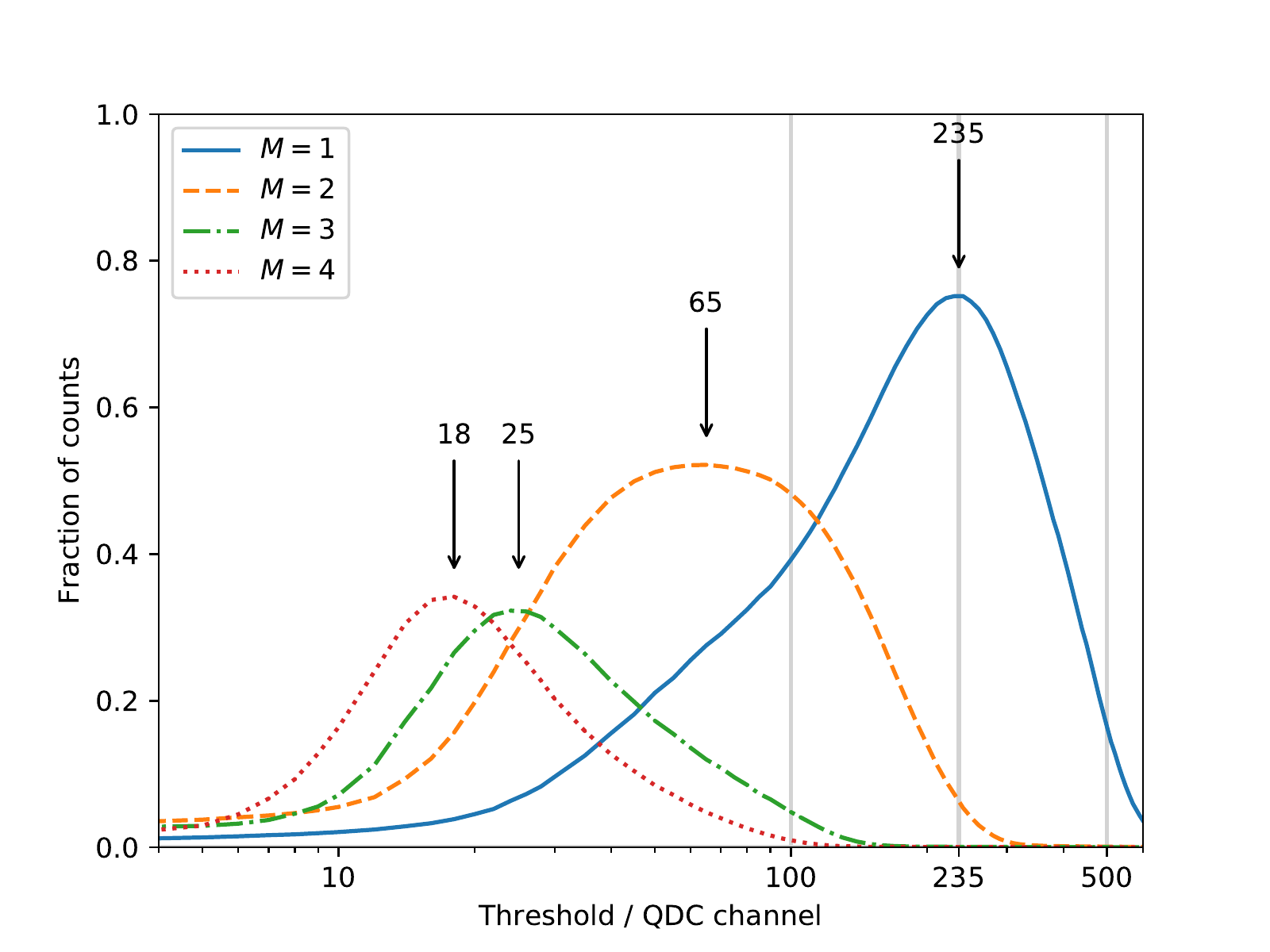}
                        \caption{Practical multiplicities}
                \label{subfig:m1to4}
                \end{subfigure}
                \caption{Tuning the analysis using a single QDC threshold. 
		Relationships between the relative number of events and
		threshold. Top Panel: $M$~$=$~0 (black dot-dot-dashed line) 
		and $M$~$>$~4 (light blue dot-dashed line). 
		Bottom Panel: $M$~$=$~1 (dark blue solid line), $M$~$=$~2 
		(orange dashed line), $M$~$=$~3 (green dot-dashed line), and 
		$M$~$=$~4 (red dotted line). The grey vertical lines at 
		QDC~channel~100, 235, and 500 represent three of the QDC 
		threshold cuts employed in Fig.~\ref{figure:contour_maps_M_with_cuts}.
		Arrows indicate the optimal values for QDC threshold cuts for
		tuning the resulting data set for a specific value of $M$.
		(For interpretation of the references to color in this 
		figure caption, the reader is referred to the web version 
		of this article.)}
		\label{figure:active_area_per_threshold}
        \end{center}
\end{figure}

The multiplicity of pixels with a signal above discriminator threshold (the
hit multiplicity denoted $M$~$=$~1, $M$~$=$~2, etc.) has been investigated as a 
function of the scintillation position and also the discrimination level. 
Figure~\ref{figure:contour_maps_M_with_cuts} displays regions in the vicinity of 
P37 where $M$~$=$~1, 2, and 3 predominate. Hits have been determined according
to the pulse heights (Fig.~\ref{figure:mean_yields_with_gain_correction}) 
exceeding discrimination levels of 100 (\ref{subfig:100_channels}), 235 
(\ref{subfig:235_channels}), and 500 (\ref{subfig:500_channels}) QDC~channels. At the 100-channel
threshold, $M$~$=$~1 events are confined to the center of a pixel. At the edges,
events are predominantly $M$~$=$~2, while in the corners, $M$~$=$~3. To see any considerable 
$M$~$=$~4 contributions around the pixel corners, lower thresholds are required
as seen in Fig.~\ref{figure:active_area_per_threshold}. Raising the threshold to 
235~channels extinguishes $M$~$=$~2 and $M$~$=$~3 almost completely. Raising even 
further to 500~channels serves merely to reduce the number of $M$~$=$~1 events.
The threshold level obviously affects the relative efficiency with which the 
SoNDe detector prototype registers $M$~$=$~1, $M$~$=$~2, etc. events, and 
Fig.~\ref{figure:contour_maps_M_with_cuts} clearly shows that there is an optimum
threshold value to maximize the number of $M$~$=$~1 events detected and also to 
maximze the area of the detector where the $M$~$=$~1 efficiency is high.

Figure~\ref{figure:active_area_per_threshold} illustrates the trend in $M$ as a 
function of QDC threshold cut when a series of 36 $\alpha$-particle beam 
measurements were performed in a 6~$\times$~6 grid covering 
the face of P37. Variable thresholds have been applied to the data, but in each case, 
the same cut has been applied to all pixels. Two ``extreme" curves are shown in 
\ref{subfig:m_extreme}: the $M$~$=$~0 curve and the $M$~$>$~4 curve. The $M$~$=$~0 curve 
corresponds to events which do not exceed the applied threshold in any pixel. 
$M$~$=$~0 events start to register at a threshold of $\sim$30~channels and rise 
steeply after channel~$\sim$200 to $\sim$95\% at channel~600. The $M$~$>$~4 curve 
corresponds to events which exceed the applied threshold in four or more pixels. 
At a threshold of $\sim$4~channels, $\sim$90\% of events are $M$~$>$~4, falling
essentially to zero at channel~$\sim$50. Four other curves are shown in 
\ref{subfig:m1to4}: $M$~$=$~1, $M$~$=$~2, $M$~$=$~3, and $M$~$=$~4, corresponding 
to events which exceed the applied threshold in one, two, three, and 
four pixels, respectively. Each of these curves demonstrate clear maxima so that 
the analysis procedure may be ``tuned" to select an event multiplicity by applying 
the appropriate threshold. The detection efficiency for $M$~$=$~1 events peaks at 
$\sim$75\% at a threshold of 235~channels, where $M$~$=$~2, 3, and 4 have negligible 
efficiency as they peak at channels~65, 25, and 18, respectively.

\section{Summary and Discussion}
\label{section:summary}

The position-dependent response of a SoNDe detector prototype, which consists 
of a 1~mm thick
wafer of GS20 scintillating glass read out by an 8~$\times$~8~pixel type H12700
MAPMT has been measured using a collimated $^{241}$Am source. The spreading of 
the scintillation light and the resulting distributions of charge on the MAPMT 
anodes were studied as a function of $\alpha$-particle interaction position by 
scanning the collimated $\alpha$-particle beam across the face of the MAPMT using 
a high precision XY coordinate translator.

Initially, pixel gain non-uniformity across the 64 MAPMT anodes was measured 
using the 3~mm collimated source positioned at each pixel center, which produced 
uniform illumination of the pixel centers.
The results, which differ from relative gain 
data provided by the MAPMT manufacturer on the 10\% level (Fig.~\ref{figure:gainmaps}), were used 
to correct all subsequent 1~mm scan data.

Anode charge distributions collected from each MAPMT pixel at each scanned 
coordinate show a strong position dependence of the signal amplitude. The single
pixel signal is strongest when the source is located at the pixel
center, and falls away as the pixel boundaries are approached 
(Fig.~\ref{figure:mean_yields_with_gain_correction}). At the pixel center, the 
signal tends to be concentrated in that pixel, while at pixel boundaries, the 
signal is shared between the adjacent pixels.

Rate and data-volume considerations for operation of SoNDe modules at ESS 
will require a relatively simple mode of operation for the SoNDe 
data-acquisition system. It will not be possible to read out multiple pixels to 
construct a weighted-mean interaction position as in an Anger Camera. Instead, 
it will be necessary to identify the pixel where the maximum charge occurs and 
record only the identity (P1 -- P64) of that pixel. 
To this end, we studied the 
effect of signal amplitude thresholds on the multiplicity of pixel hits (that 
is, the number of signals above threshold) as a function of the $\alpha$-particle 
interaction position (Fig.~\ref{figure:contour_maps_M_with_cuts}). 
This study showed that there is an optimum discrimination level which 
maximizes the number of single-pixel or $M$~$=$~1 hits. Below this level, 
multi-pixel hits start to dominate, while above this level, the single-pixel
efficiency drops (Fig.~\ref{figure:active_area_per_threshold}). At the optimum 
discrimination level, which under the present operating conditions was 
channel~235, $\sim$75\% of the $\alpha$-particle interactions were registered 
as single pixel.

Further work pertaining to the characterization of the SoNDe detector prototype 
is progressing in parallel to the project reported here. This includes the 
development of a simulation within the \geant~framework to complement and 
extend the ray-tracing simulation mentioned in this work. This \geant~model 
fully simulates the interactions of ionizing radiation in the GS20 and tracks 
the produced scintillation photons to the MAPMT cathode. It is being used to 
study the effects of optical coupling, surface finish, and partial pixelation 
(by machining grooves) of the GS20 wafer on the response of the detector. The 
relative response of a SoNDe detector prototype to different incident particle 
species is also being studied. Irradiation of a SoNDe detector prototype with 
a collimated beam of thermal neutrons has been performed and data analysis is 
nearing completion. An extension of these studies, where an optical diffuser is 
inserted between the GS20 and the MAPMT, is planned to examine a possible 
Anger camera mode of operation. And further position-dependence studies will 
be made using fine needle-like beams of a few MeV protons and deuterons produced 
by an accelerator. This will allow determination of the scintillation signal 
as a function of interacting particle species and will be complemented by work 
with fast-neutron and gamma-ray sources.

\section*{Acknowledgements}
\label{acknowledgements}

We thank Prof. David Sanderson from the Scottish Universities Environmental 
Research Centre for providing $\alpha$-spectroscopy facilities for the 
calibration of our $^{241}$Am source. We acknowledge the support of 
the European Union via the Horizon 2020 Solid-State Neutron Detector Project, 
Proposal ID 654124, and the BrightnESS Project, Proposal ID 676548. We also
acknowledge the support of the UK Science and Technology Facilities Council 
(Grant nos. STFC 57071/1 and STFC 50727/1) and UK Engineering and Physical 
Sciences Research Council Centre for Doctoral Training in Intelligent 
Sensing and Measurement (Grant No. EP/L016753/1).

\newpage

\bibliographystyle{elsarticle-num}

\end{document}